\documentclass[12pt]{article}
\usepackage{amsmath}
\def\be{\begin{equation}}
\def\ee{\end{equation}}
\def\arr{\begin{array}{rll}}
\def\ea{\end{array}}
\def\bea{\begin{eqnarray}}
\def\eea{\end{eqnarray}}

\def\N2{$N{=}2$}

\def\>{\rangle}
\def\<{\langle}
\def\+{\dagger}
\def\={\ =\ }

\begin{document}
\renewcommand{\thefootnote}{\fnsymbol{footnote}}
\begin{titlepage}
\setcounter{page}{0}
\vskip 1cm
\begin{center}
{\LARGE\bf Ricci--flat spacetimes with  }\\
\vskip 0.5cm
{\LARGE\bf $l$--conformal Galilei symmetry}\\
\vskip 1cm
$
\textrm{\Large D. Chernyavsky and A. Galajinsky \ }
$
\vskip 0.7cm
{\it
Laboratory of Mathematical Physics, Tomsk Polytechnic University, \\
634050 Tomsk, Lenin Ave. 30, Russian Federation} \\
{E-mails: chernyavsky@tpu.ru, galajin@tpu.ru}

\end{center}
\vskip 1cm
\begin{abstract} \noindent
Ricci--flat metrics of the ultrahyperbolic signature which enjoy the $l$--conformal Galilei symmetry are constructed. They involve
the $AdS_2$--metric in a way similar to the near horizon black hole geometries. The associated geodesic equations are shown to describe a second order
dynamical system for which the acceleration generators are functionally independent.
\end{abstract}


\vskip 0.5cm


\end{titlepage}

\renewcommand{\thefootnote}{\arabic{footnote}}
\setcounter{footnote}0

\noindent
{\bf 1. Introduction}\\

\noindent
Motivated by the ongoing studies of the nonrelativistic AdS/CFT--correspondence, dynamical realizations of the $l$--conformal Galilei/Newton--Hooke algebra \cite{Barut}--\cite{nor} have been attracting increasing
interest \cite{Stichel}--\cite{KA} \footnote{Note that the $l$-conformal Galilei and Newton--Hooke algebras are
isomorphic (see e.g. \cite{GM}). However, as far as dynamical realizations are concerned, a linear change of the basis which relates them implies a change of the Hamiltonian
which alters the dynamics.}. A peculiar feature of the algebra is that along with
the generators of time translation, dilatation, special conformal transformation, spatial rotations,
spatial translations, and Galilei boosts, it also includes accelerations. All together,
there are $2l+1$ vector generators, where $l$ is a (half)integer number. The latter enters the structure relations of the algebra and gives rise to the term "$l$--conformal" \cite{nor}.

In spite of the extensive studies in \cite{Stichel}--\cite{KA}, there are several issues concerning the dynamical realizations constructed so far. Firstly, all examples for which the acceleration generators are functionally independent involve higher derivative terms. In this regard it is worth mentioning that interacting higher derivative systems typically show up instability in classical dynamics and bring about violation of unitarity and/or trouble with ghosts in quantum theory. Secondly, consistent second order systems invariant under the action of the $l$--conformal Galilei/Newton--Hooke group have been constructed in \cite{GM3,GM1}. However, the acceleration generators prove to be functionally dependent. The first two points raise the question of whether a second order
dynamical system exists for which the acceleration generators are functionally independent.
Thirdly, applications of the $l$--conformal Galilei/Newton--Hooke symmetry within the general relativistic context are unknown but for the trivial instance of $l=\frac 12$ for which the accelerations are absent.

The goal of this work is to dwell on the last two issues. Below we construct Ricci--flat metrics of the ultrahyperbolic signature which enjoy the $l$--conformal Galilei isometry group.
A salient feature of such metrics is that they involve the $AdS_2$--metric in a way similar to the near horizon black hole geometries (see, e.g., \cite{AG} and references therein). Hence, they may find potential applications within the context of the AdS/CFT--correspondence. We also argue that the geodesic equations on such a spacetime provide the first example in the literature of a consistent second order dynamical system which accommodates the $l$--conformal Galilei symmetry in such a way that the acceleration generators are functionally independent.

The work is organized as follows. In Sect. 2 we briefly outline the generic group theoretic construction and build the $AdS_2$--metric starting from the $so(2,1)$--subalgebra in the $l$--conformal Galilei algebra. Sect. 3 extends the analysis to the full algebra. In particular, the Maurer--Cartan one--forms, which transform homogeneously under the dilatations, special conformal transformations and rotations, while
hold invariant with respect to the remaining group transformations, are constructed. These forms are used in Sect. 4 with the aid to construct novel Ricci--flat metrics with the $l$--conformal Galilei isometry group. In Sect. 5 we consider the geodesic equations and argue that they describe a second order dynamical system for which the acceleration generators are functionally independent. The concluding Sect. 6 contains the summary and the discussion of possible further developments.

\vspace{0.5cm}

\noindent
{\bf 2. $AdS_2$--metric from $so(2,1)$}\\

\noindent
As is well known, the metric of the two--dimensional anti de Sitter spacetime naturally arises if one applies the conventional group theoretic construction to the conformal algebra in one dimension $so(2,1)$. Since in the next sections we will use the same method for building a metric invariant under the action of the $l$--conformal Galilei group, for which $SO(2,1)$ is a subgroup, below we briefly remind the reader the derivation of the $AdS_2$--metric from the $so(2,1)$ algebra. For simplicity, throughout the text we set the radius of curvature of $AdS_2$ to be the unity.

One starts with the structure relations of $so(2,1)$
\be\label{so}
[H,D]=i H, \qquad [H,K]=2 i D, \qquad [D,K]=i K,
\ee
where $H$, $D$, and $K$ are the generators of time translations, dilatations, and special conformal transformations, respectively. Then within the corresponding conformal group $G=SO(2,1)$ one chooses the subgroup $L$ generated by $D$ and considers the coset space $G/L$
\be
\tilde G= e^{itH} e^{irK} \times L,
\ee
which is parametrized by two real coordinates $t$ and $r$.

Left multiplication by the group element
$g=e^{iaH} e^{ibK} e^{icD}$
determines the action of the group on the coset
\be\label{ctr}
t'=t+a+b t^2+ct, \qquad \quad r'=r+b(1-2tr)-cr,
\ee
where the Baker--Campbell--Hausdorff formula
\be
e^{iA}~ T~ e^{-iA}=T+\sum_{n=1}^\infty\frac{i^n}{n!}
\underbrace{[A,[A, \dots [A,T] \dots]}_{n~\rm times}
\ee
has been used.

The Maurer--Cartan one--forms ${\tilde G}^{-1} d \tilde G=i(\omega_H H+\omega_K K+\omega_D D)$, where
\be
\omega_H=dt, \quad \omega_K=r^2 dt+dr, \quad \omega_D=-2r dt,
\ee
are the building blocks for constructing a metric invariant under the action of $SO(2,1)$. Considering a generic quadratic form constructed out of $\omega_H$, $\omega_K$, $\omega_D$, and demanding that it
be invariant under the conformal transformations (\ref{ctr}), one gets (up to an irrelevant constant factor)
\be\label{1}
ds^2=\omega_H \omega_K=(r^2 dt^2+dt dr).
\ee
The redefinition of the temporal coordinate
\be\label{ttr}
t=\frac 12 \left(\tilde t+\frac{1}{r} \right)
\ee
brings (\ref{1}) to the conventional $AdS_2$--metric written in the Poincar\'e coordinates
\be
ds^2=r^2 d\tilde t {}^2-\frac{dr^2}{r^2},
\ee
where the overall factor $\frac 14$ has been discarded. The transformation (\ref{ttr}) keeps the form of the time translation and dilatation intact
\be
\tilde t'=\tilde t+a+c\tilde t, \qquad \quad r'=r-cr,
\ee
while the special conformal transformation now reads
\be
{\tilde t}'=\tilde t+\frac{b}{2} \left({\tilde t}^2+\frac{1}{r^2} \right), \qquad r'=r-b r \tilde t.
\ee

Note that from the very beginning one could consider the three--dimensional group manifold $\tilde G= e^{itH} e^{irK} e^{i u D}$ and construct a metric invariant under the action of the $SO(2,1)$--group out of the left--invariant Maurer--Cartan one--forms. It is straightforward to verify that it does not solve the vacuum Einstein equations, which forces one to choose the lower dimensional coset space as described above.

\vspace{0.5cm}

\noindent
{\bf 3. Maurer--Cartan one--forms for the $l$--conformal Galilei algebra}\\

\noindent

A peculiar feature of the $l$--conformal Galilei algebra is that along with
the generators of time translation $H$, dilatation $D$, special conformal transformation $K$, spatial rotations $M_{ij}$ (with $i=1,\dots,d$),
spatial translations $C^{(0)}_i$, and Galilei boosts $C^{(1)}_i$ it also includes the generators of
accelerations $C^{(\alpha)}_i$, where $\alpha=2,\dots, 2l$ and $l$ is a (half)integer number. Combining the vector generators into a single object $C^{(n)}_i$, where $n=0,\dots, 2l$, one gets the structure relations of the algebra \cite{nor} which involve (\ref{so}) along with
\bea\label{algebra}
&&
[H,C^{(n)}_i]=i n C^{(n-1)}_i, \qquad
[D,C^{(n)}_i]=i (n-l) C^{(n)}_i, \qquad [K,C^{(n)}_i]=i (n-2l) C^{(n+1)}_i,
\\[2pt]
&&
[M_{ij},C^{(n)}_k]=-i (\delta_{ik} C^{(n)}_j-\delta_{jk} C^{(n)}_i), ~ [M_{ij},M_{kl}]=-i(\delta_{ik} M_{jl}+\delta_{jl} M_{ik}-
\delta_{il} M_{jk}-\delta_{jk} M_{il}).
\nonumber
\eea

In order to construct a metric invariant under the action of the $l$--conformal Galilei group,
we choose a subgroup $L$ generated by $D$ and $M_{ij}$ and consider the coset space
\be
\tilde G=e^{itH} e^{irK} e^{i x^{(n)}_i C^{(n)}_i} \times L
\ee
parametrized by the coordinates $t$, $r$ and $x^{(n)}_i$. As usual, summation over repeated indices is understood unless otherwise is explicitly stated. Left multiplication by the group element
$g=e^{iaH} e^{ibK} e^{icD} e^{i \lambda^{(n)}_i C^{(n)}_i} e^{\frac{i}{2} \omega_{ij} M_{ij}}$
determines the action of the group on the coset
\bea\label{transf}
&&
t'=t+a+b t^2+ct, \qquad \quad r'=r+b(1-2tr)-cr,
\nonumber\\[2pt]
&&
x'^{(n)}_i=x^{(n)}_i-2bt(n-l)x^{(n)}_i-c(n-l)x^{(n)}_i-\sigma_{ij} x^{(n)}_j+
\nonumber\\[4pt]
&&
\qquad \quad
+\sum_{s=0}^n \sum_{m=s}^{2l}\frac{{(-1)}^{n-s} m! (2l-s)!}{s! (m-s)! (n-s)! (2l-n)!} t^{m-s} r^{n-s} \lambda^{(m)}_i,
\eea
where $a,b,c,\lambda^{(n)}_i$ and $\sigma_{ij}=-\sigma_{ji}$
are infinitesimal parameters corresponding to the time translations, special conformal transformations, dilatations, vector generators in the algebra,
and spatial rotations, respectively.

The Maurer-Cartan one--forms ${\tilde G}^{-1} d \tilde G=i(\omega_H H+\omega_K K+\omega_D D+\omega^{(n)}_i C^{(n)}_i)$
\bea\label{MC}
&&
\omega^{(n)}_i=d x^{(n)}_i+2r(n-l) x^{(n)}_i dt-(n+1)  x^{(n+1)}_i dt-(n-2l-1)  x^{(n-1)}_i (r^2 dt+dr),
\nonumber\\[4pt]
&&
\omega_H=dt, \qquad \qquad  \omega_K=r^2 dt+dr, \qquad \qquad  \omega_D=-2r dt,
\eea
where we have set
\be
x^{(-1)}_i=x^{(2l+1)}_i=0
\ee
provide the building blocks for constructing a metric with the $l$--conformal Galilei isometry group. Bearing in mind potential applications within the  context of the AdS/CFT--correspondence,  we choose to switch to the new temporal coordinate (\ref{ttr}) which brings the metric (\ref{1}) to the conventional $AdS_2$ form and changes $\omega^{(n)}_i$ as follows:
\bea
&&
\tilde\omega^{(n)}_i=d x^{(n)}_i+\left( r(n-l) x^{(n)}_i -\frac 12 (n+1)  x^{(n+1)}_i \right) \left(d\tilde t-\frac{dr}{r^2}\right)-
\nonumber\\[2pt]
&&
\qquad \quad -\frac 12 (n-2l-1)  x^{(n-1)}_i r^2 \left(d\tilde t+\frac{dr}{r^2}\right).
\eea

\vspace{0.5cm}

\noindent
{\bf 4. Ricci--flat metrics with the $l$--conformal Galilei isometry group}\\

\noindent
Taking into account the fact that $\tilde\omega^{(n)}_i$ are invariant under the time translation and $\lambda^{(n)}_i$--transformations, while with respect to the dilatations, special conformal transformations and rotations they transform homogeneously
\bea
&&
\tilde\omega'^{(n)}_i=(1-c(n-l))\tilde\omega^{(n)}_i, \qquad \tilde\omega'^{(n)}_i=\left(1-b(n-l)\left(\tilde t+\frac{1}{r} \right)\right)\tilde\omega^{(n)}_i,
\nonumber\\[2pt]
&&
\tilde\omega'^{(n)}_i=(\delta_{ij}-\sigma_{ij}) \tilde\omega^{(n)}_j
\eea
one can immediately construct a metric which holds invariant under the action of the $l$--conformal Galilei group
\bea\label{metric}
&&
ds^2=\left(r^2 d\tilde t {}^2-\frac{dr^2}{r^2}\right)+S_{n,m} \tilde\omega^{(n)}_i \tilde\omega^{(m)}_i,
\eea
where $S_{n,m}$ is an off--diagonal symmetric matrix (no sum with respect to the index $n$ on the right hand side)
\be\label{S}
S_{n,m}=S_{m,n}=k_n  \delta_{n+m,2l},
\ee
which involves constant entries $k_n$ to be fixed below. Note that for the case of $d=1$, $l=1$ conformal Galilei algebra
similar metrics have been studied in \cite{BK}.

A straightforward computation shows that (\ref{metric}) is not Ricci--flat. In order to promote it to a solution of the vacuum Einstein equations, let us minimally extend the metric by an extra coordinate $y$
\be\label{metric1}
ds^2=\alpha(y)\left(r^2 d\tilde t {}^2-\frac{dr^2}{r^2}\right)+S_{n,m} \tilde\omega^{(n)}_i \tilde\omega^{(m)}_i+\epsilon dy^2,
\ee
where $\alpha(y)$ is a function to be fixed below and  $\epsilon=\pm 1$. It is assumed that $y$ remains intact under the $l$--conformal Galilei transformations so that the metric maintains the symmetry of its predecessor  (\ref{metric}).
One could promote $k_n$ in (\ref{S}) to become real functions $k_n(y)$ as well. However, when solving the Einstein equations, these lead to a nonlinear system of ordinary differential equations which, unfortunately, we fail to solve. Note that an arbitrary function of $y$ in front of $dy^2$ can always be eliminated by redefining $y$, leaving one with the sign choice arbitrariness expressed by $\epsilon=\pm 1$.

The analysis of the Einstein equations is greatly facilitated if one introduces the notation
\be
\tilde\omega^{(n)}_i=d x^{(n)}_i+ a^{(n)}_i d\tilde t+b^{(n)}_i dr,
\ee
where
\bea\label{ab}
&&
a^{(n)}_i=r(n-l) x^{(n)}_i -\frac 12 (n+1)  x^{(n+1)}_i-\frac 12 r^2 (n-2l-1)  x^{(n-1)}_i,
\nonumber\\[2pt]
&&
b^{(n)}_i=-\frac{1}{r} (n-l) x^{(n)}_i +\frac{1}{2r^2} (n+1)  x^{(n+1)}_i
-\frac 12 (n-2l-1)  x^{(n-1)}_i,
\eea
and rewrites the metric, its inverse, and the Christoffel symbols in terms of the vectors $a^{(n)}_i$ and $b^{(n)}_i$. From the equation $R_{yy}=0$ one gets
\be
{\left(\frac{\alpha'}{\alpha}\right)}'+\frac 12 {\left(\frac{\alpha'}{\alpha}\right)}^2=0,
\ee
which implies that $\alpha(y)$ is a quadratic function of its argument
\be\label{al}
\alpha(y)=c_1{(y+c_2)}^2,
\ee
where $c_1$ and $c_2$ are arbitrary constants. The latter can always be eliminated by the redefinition $y~\rightarrow y+c_2$. In what follows we will disregard it.

The components of the Ricci tensor $R_{ty}$, $R_{ry}$, $R_{x y}$ prove to vanish identically, provided one takes into account that $a^{(n)}_i$ and $b^{(n)}_i$ in (\ref{ab}) are divergence free
\be
\frac{\partial a^{(n)}_i}{\partial x^{(n)}_i}=0, \qquad \frac{\partial b^{(n)}_i}{\partial x^{(n)}_i}=0.
\ee

The equation $R_{xx}=0$ yields the conditions which allow one to fix the components of the matrix $S_{n,m}$
\bea\label{const1}
&&
n^2 S_{n-1,2l-n+1}+{(2l-n)}^2 S_{n+1,2l-n-1}
\nonumber\\[2pt]
&&
-\left[ {(n+1)}^2 {\tilde S}^{n+1,2l-n-1}+{(n-2l-1)}^2 {\tilde S}^{n-1,2l-n+1} \right] {\left(S_{n,2l-n}\right)}^2=0,
\eea
where $n=0,\dots,2l$. In Eq. (\ref{const1}) ${\tilde S}^{n,m}={\tilde S}^{m,n}=\frac{1}{S_{n,m}}=\frac{1}{k_n}  \delta_{n+m,2l}$ (no sum in $n$) stands for the inverse of $S_{n,m}$ and
it is assumed that
\be
S_{2l+1,n}=S_{-1,n}={\tilde S}^{2l+1,n}={\tilde S}^{-1,n}=0.
\ee
It is straightforward to verify that Eqs. (\ref{const1}) algebraically relate all the components $S_{n,m}$ to $S_{0,2l}$, the latter being unspecified. For the absolute values one reveals the recurrence
\be
|S_{n+1,2l-n-1}|=\frac{n+1}{2l-n}|S_{n,2l-n}|
\ee
with $n=0,\dots,2l-1$. Because the flip of sign is admissible, the total number of options grows with $l$. Some examples are given below.

In a similar fashion $R_{tt}=0$ links the constant $c_1=\frac 12 \alpha''$ in (\ref{al}) to the parameters $d$, $l$, $\epsilon$ and the matrix elements $S_{p,q}$, ${\tilde S}^{p,q}$
\be\label{const2}
c_1=\epsilon+\frac{ l(l+1)(2l+1) d \epsilon}{6}-\frac{d \epsilon}{4} \sum_{p=0}^{2l-1} \sum_{q=1}^{2l} (p+1)(q-2l-1) {\tilde S}^{p+1,q-1} S_{p,q}.
\ee
All other components of the Ricci tensor prove to vanish provided (\ref{const1}) and (\ref{const2}) hold.

Given $d$ and $l$, the equations (\ref{const1}) and (\ref{const2}) are readily solved. In particular, for $l=\frac 12, 1,\frac 32,2$ one finds
\begin{align}
&
l=\frac 12;  && &&  && c_1=\epsilon \left(1+\frac{d}{2} \right);
\nonumber\\[2pt]
&
l=1; && S_{1,1}=\frac 12 S_{0,2} && && c_1=\epsilon\left(1+2d\right)
\nonumber\\[2pt]
&
&&
S_{1,1}=-\frac 12 S_{0,2} &&  && c_1=\epsilon;
\nonumber\\[2pt]
&
l=\frac 32; &&
S_{1,2}=\frac 13 S_{0,3} && && c_1=\epsilon\left(1+5d \right)
\nonumber\\[2pt]
&
&& S_{1,2}=-\frac 13 S_{0,3} && && c_1=\epsilon\left(1+2d \right);
\nonumber\\[2pt]
&
l=2; && S_{2,2}=\frac 16 S_{0,4}, && S_{1,3}=\frac 14 S_{0,4} && c_1=\epsilon\left(1+10d \right)
\nonumber\\[2pt]
&
&& S_{2,2}=-\frac 16 S_{0,4}, && S_{1,3}=\frac 14 S_{0,4} && c_1=\epsilon\left(1+4d \right)
\nonumber\\[2pt]
&
&&S_{2,2}=-\frac 16 S_{0,4}, && S_{1,3}=-\frac 14 S_{0,4} && c_1=\epsilon\left(1+6d \right)
\nonumber\\[2pt]
&
&& S_{2,2}=\frac 16 S_{0,4}, && S_{1,3}=-\frac 14 S_{0,4} && c_1=\epsilon.
\nonumber
\end{align}
It is to be remembered that in each case $S_{0,2l}$ remains unspecified.

Having fixed $S_{n,m}$, let us comment on the signature of a spacetime endowed with the metric (\ref{metric1}).
Given the spatial dimension $d$ in which the original $l$--conformal Galilei algebra (\ref{algebra}) is realized and the value of the (half)integer parameter $l$, the $AdS_2$--spacetime in (\ref{metric1}) is extended by $(2l+1)d$ extra dimensions parametrized by the coordinates $x^{(0)}_i,\dots,x^{(2l)}_i$, $i=1,\dots,d$. For half--integer $l$, these are split into $\frac{(2l+1)d}{2}$ spatial and $\frac{(2l+1)d}{2}$ temporal dimensions, while for integer $l$ one reveals $ld$ temporal and $(l+1)d$ spatial dimensions or vice versa depending on which sign is chosen for the components $S_{n,m}$ linked to $S_{0,2l}$. Depending on whether $\epsilon=1$ or $\epsilon=-1$ is chosen in the last term in (\ref{metric1}), the remaining coordinate $y$ brings about one more temporal or spatial dimension. Assuming that $y$ has the dimension of length, the coordinates $r$, $t$ and the parameter $S_{0,2l}$ should be considered dimensionless\footnote{The dimensionless coordinates $r$ and $t$ are linked to $(r',t')$ parametrizing $AdS_2$ in the obvious way $r=\frac{r'}{R}$, $ct=\frac{ct'}{R}$, where $R$ is the radius of curvature of $AdS_2$ and $c$ is the speed of light which throughout the text we set to be the unity.}, while $x^{(n)}_i$ must have the dimension of length.

To summarize, (\ref{metric1}) describes a $[(2l+1)d+3]$--dimensional Ricci--flat spacetime of the ultrahyperbolic signature which is characterized by one free parameter $S_{0,2l}$.

\vspace{0.5cm}

\noindent
{\bf 5.  Geodesics as the dynamical realizations of the $l$--conformal Galilei group}\\

\noindent
The metrics constructed in the previous section allow one to immediately obtain novel dynamical realizations of the $l$--conformal Galilei group. It suffices to consider the geodesic equations
\be\label{geo}
\frac{d^2 Z^A}{d \tau^2}+\Gamma^{A}_{BC} (Z) \frac{d Z^B}{d \tau}  \frac{d Z^C}{d \tau}=0,
\ee
where $Z^A=(t,r,x^{(0)}_i,\dots,x^{(2l)}_i,y)$, with $i=1,\dots,d$, and $\tau$ is the proper time. As was mentioned in the Introduction,
apart from the oscillator--like models for which the acceleration generators are functionally dependent \cite{GM3,GM1}, all other dynamical realizations known in the literature prove to involve higher derivative terms.
That (\ref{geo}) is a genuine second order system is argued by appealing to the Riemann coordinates. As is well known (see, e.g. \cite{Eis}), in the normal neighborhood of a given point $Z^A_0$ on a manifold one can introduce a change of the coordinate system $Z^A~\rightarrow~Y^A(Z)$ such that $\Gamma'^A_{BC}(Y_0)=0$, where $Y_0^A=Y^A(Z_0)$, and the geodesic passing through $Y^A_0$ is the straight line
\be\label{geo1}
\frac{d^2 Y^A}{d \tau^2}=0,
\ee
which, obviously, is a second order system.

Another way of arguing is to put the system (\ref{geo}) into the conventional Hamiltonian form. Introducing momenta $P_A$ canonically conjugate to the configuration space variables $Z^A$, the standard Poisson bracket $\{Z^A,P_B\}={\delta^A}_B$, and the Hamiltonian
\be\label{hami}
\mathcal{H}=\frac 12 g^{AB} (Z) P_A P_B,
\ee
where $g^{AB}(Z)$ is the inverse metric,
one can readily check that Hamilton's equations of motion following from (\ref{hami}) are equivalent to (\ref{geo}). The fact that (\ref{hami}) does not involve unconventional kinetic terms linear in momenta for some of the degrees of freedom
implies that it is not of Ostrogradsky's type.

Thus, Eqs. (\ref{geo}) represent a second order dynamical system which holds invariant under the action of the $l$--conformal Galilei group. Its integrals of motion can be constructed from the Killing vector fields $\xi^A(Z) \frac{\partial}{\partial Z^A}$ associated with the metric (\ref{metric1}) in the usual way: $\xi^A(Z) g_{AB} (Z) \frac{d Z^B}{d \tau}$.
The explicit form of $\xi^A(Z)$ is readily derived from Eq. (\ref{transf}) by taking into account the redefinition of the temporal coordinate (\ref{ttr}).

As an example, let us consider the Killing vector fields for the case of $l=1$
\bea
&&
H=\frac{\partial}{\partial \tilde t}, \qquad  D=\tilde t \frac{\partial}{\partial \tilde t}-r \frac{\partial}{\partial r}+x^{(0)}_i \frac{\partial}{\partial x^{(0)}_i}-x^{(2)}_i \frac{\partial}{\partial x^{(2)}_i},
\nonumber\\[2pt]
&&
K=\frac 12 \left({\tilde t}^2+\frac{1}{r^2} \right) \frac{\partial}{\partial \tilde t}-r \tilde t \frac{\partial}{\partial r}+\left(\tilde t+\frac{1}{r} \right) x^{(0)}_i \frac{\partial}{\partial x^{(0)}_i}-
\left(\tilde t+\frac{1}{r} \right) x^{(2)}_i \frac{\partial}{\partial x^{(2)}_i},
\nonumber\\[2pt]
&&
C^{(0)}_i=\frac{\partial}{\partial x^{(0)}_i}-2 r\frac{\partial}{\partial x^{(1)}_i}+r^2 \frac{\partial}{\partial x^{(2)}_i},
\nonumber\\[2pt]
&&
C^{(1)}_i=\frac 12 \left(\tilde t+\frac{1}{r} \right) \frac{\partial}{\partial x^{(0)}_i} -r \tilde t \frac{\partial}{\partial x^{(1)}_i}+\frac{r^2}{2} \left(\tilde t-\frac{1}{r} \right) \frac{\partial}{\partial x^{(2)}_i},
\nonumber\\[2pt]
&&
C^{(2)}_i=\frac 14 {\left(\tilde t +\frac{1}{r} \right)}^2 \frac{\partial}{\partial x^{(0)}_i}+\frac{r}{2} \left(\frac{1}{r^2}-{\tilde t}^2 \right) \frac{\partial}{\partial x^{(1)}_i}+\frac{r^2}{4} {\left(\tilde t-\frac{1}{r} \right)}^2 \frac{\partial}{\partial x^{(2)}_i}.
\eea
It is straightforward to verify that the corresponding integrals of motion are functionally independent. As usual, it suffices to consider the gradients of the first integrals and to verify that they are linearly independent.
Higher values of $l$ are treated likewise.

Thus, the geodesic equations constructed with the aid of the metric (\ref{metric1}) provide the first example of a second order dynamical system for which the acceleration generators are functionally independent. As compared to the formulations previously studied in the literature, a specific novel feature is that to each acceleration generator $C^{(n)}_i$ from the $l$--conformal Galilei algebra there correspond extra spacetime dimensions parametrized by the coordinates $x^{(n)}_i$. To put it in other words, the dimension of spacetime, in which the geodesic equations are formulated, grows with $l$. This allows one to accommodate the full $l$--conformal Galilei symmetry in the geodesic equations thus avoiding the unpleasant features like the presence of higher derivative terms or the functional dependence of generators of symmetry transformations.

\vspace{0.5cm}

\noindent
{\bf 6. Conclusion}\\

\noindent
To summarize, in this work we applied the method of nonlinear realizations so as to construct novel solutions to the vacuum Einstein equations which enjoy the $l$--conformal Galilei isometry group. Given the values of the parameter $l$ and
the spatial dimension $d$, in which the original $l$--conformal Galilei algebra is realized, the method yields a $[(2l+1)d+3]$--dimensional spacetime of the ultrahyperbolic signature. Geodesic equations defined on such a
spacetime automatically
inherit symmetries of the background metric and thus describe a second order
dynamical system for which the acceleration generators are functionally independent.

Turning to possible further developments, the most interesting issue is to study whether the metrics constructed in this work can be used within the context of the "Near horizon black hole/CFT" correspondence (for a review see, e.g., \cite{GC}). While their structure very much resembles the near horizon black hole geometries, the signature is ultrahyperbolic so some subtleties may occur. It is also worth studying whether the metrics in this work can be obtained as the near horizon limit of more general solutions of the vacuum Einstein equations which enjoy the acceleration enlarged Galilei symmetry but does not possess the $so(2,1)$ conformal invariance.  As the number of the first integrals of the geodesics equations proves to be sufficient to ensure integrability, it would be interesting to explicitly integrate the equations of motion. Finally, it is worth studying whether other parametrizations of the coset space may yield solutions to the vacuum Einstein equations.

\vspace{0.5cm}

\noindent{\bf Acknowledgements}\\

\noindent
The authors are grateful to Dmitry Zimin and the Dynasty Foundation for the invaluable support of the fundamental science and education in Russia in the period from 2002 till 2015.
This work was supported by the MSE program Nauka under the project 3.825.2014/K and the RFBR grant 15-52-05022.

\end{document}